\documentclass[twocolumn,showpacs,preprintnumbers,amsmath,amssymb,prl,superscriptaddress]{revtex4-1}
\usepackage{bbm}
\usepackage{mathrsfs}
\usepackage{graphicx}
\usepackage{dcolumn}
\usepackage{bm}
\usepackage{amsmath}
\usepackage{amsfonts}
\usepackage{color}
\usepackage[colorlinks=true,citecolor=blue,anchorcolor=blue]{hyperref}
\usepackage{floatrow}

\begin{document}

\title{Controllable Majorana fermions on domain walls of a magnetic topological insulator}
\author{Jing Wang}
\affiliation{State Key Laboratory of Surface Physics, Department of Physics, Fudan University, Shanghai 200433, China}
\affiliation{Institute for Nanoelectronic Devices and Quantum Computing, Fudan University, Shanghai 200433, China}
\affiliation{Collaborative Innovation Center of Advanced Microstructures, Nanjing 210093, China}

\begin{abstract}
We propose to realize a one-dimensional chiral topological superconducting state at the magnetic domain walls stripe of a magnetic topological insulator coupled with a conventional $s$-wave superconductor.  The localized Majorana zero modes can be constructed in a reconfigurable manner through magnetic domain writing by magnetic force microscopy. This proposal could be further extended to the Majorana zero modes at domain walls on superconducting spin-helical Dirac surface states, and may be applicable to the two-dimensional time-reversal invariant topological superconductor on FeTe$_{0.5}$Se$_{0.5}$ surface.
\end{abstract}

\date{\today}


\maketitle

The search for Majorana fermions in topological states of quantum matter has attracted intensive interest in condensed matter physics~\cite{hasan2010,qi2011,kitaev2003,nayak2008,alicea2012,beenakker2013}. Majorana zero modes (MZMs), the point-like zero energy Majorana fermions, have potential applications in topological quantum computation~\cite{kitaev2003,nayak2008} because of their exotic non-Abelian quantum statistics~\cite{moore1991,read2000,ivanov2001}. Several promising electronic systems hosting MZMs include $\nu=5/2$ fractional quantum Hall state~\cite{moore1991}, topological insulator-superconductor structures~\cite{fu2008,sun2017}, spin-orbit coupling (SOC) semiconductor nanowire-superconductor structures~\cite{sau2010,oreg2010,alicea2010,lutchyn2010,mourik2012,deng2016}, and ferromagnetic (FM) atomic chains on superconductors~\cite{nadj-Perge2013,braunecker2013,yazdani2014}. The signature of MZMs has been spectroscopically demonstrated as the zero-bias conductance peaks~\cite{mourik2012,yazdani2014,deng2016} and quantized peak value of $2e^2/h$~\cite{law2009,liucx2017,zhangh2018}. However, fabricating scalable Majorana qubits in semiconductor nanowires remains challenging~\cite{freedman2017}. Meanwhile, as one-dimensional (1D) cousin of MZMs, chiral Majorana fermions emerge as the gapless edge states of the 2D $p+ip$ chiral superconductors~\cite{read2000,fu2009a,akhmerov2009,tanaka2009a,qi2010b,chung2011,wang2015c,he2017}. The propagating chiral Majorana fermions could also lead to non-abelian braiding~\cite{lian2017} and may be useful in quantum computation. In particular, 2D chiral topological superconductor (TSC) with a Bogoliubov-de Gennes (BdG) Chern number $N=1$ can be realized in a heterostructure of a 2D quantum anomalous Hall (QAH) insulator FM film~\cite{qi2008,chang2013b,wang2015d} and an $s$-wave superconductor~\cite{qi2010b,chung2011,wang2015c}. A half-quantized two terminal conductance plateau of value $e^2/2h$ is observed in the experiment~\cite{he2017}, which signals the occurrence of a chiral Majorana edge fermion and the realization of an $N=1$ chiral superconductor~\cite{lian2018,huang2018,ji2018}.

The $\pi$ flux vortex of such a $N=1$ chiral superconductor carries a single MZM, however, manipulating individual vortex in a deterministic way is challenging. The dimensional reduction from a 2D chiral superconductor leads to a 1D chiral superconductor with a $\mathcal{Z}_2$ classification~\cite{schnyder2008,kitaev2009}. This motivates us to study the possible 1D TSC phases based on superconducting proximity coupled magnetic topological insulators (TIs), which may provide a new platform for braiding MZMs. The goal of this paper is to demonstrate that the 1D chiral TSC state can be realized at the magnetic domain walls (DWs) stripe of a superconducting magnetic TI, where the localized end MZMs can be constructed in a reconfigurable manner. This proposal could be further extended to DW stripe on superconducting spin-helical Dirac surface states, and may be applicable to the newly discovered time-reversal invariant TSC on FeTe$_{0.5}$Se$_{0.5}$ surface~\cite{wangzj2015,wu2016,xu2016,zhang2018}.

\emph{Basic physics.} The basic mechanism for chiral TSC is to creates a 1D system with a single pair of Fermi points from interplay of SOC and magnetization, therefore the proximity effect with an $s$-wave superconductor will induce a TSC~\cite{kitaev2001}. The SOC and FM ordering in magnetic TIs combine to give rise to the QAH state characterized by a finite Chern number $C$~\cite{thouless1982}, where gapless chiral edge states (CESs) appear at the sample edges as well as DWs. The CESs at opposite edges or DWs form an effective 1D electronic channel. Specifically, the $C=1$ QAH state has been realized in magnetic TIs of Cr- or V-doped (Bi,Sb)$_2$Te$_3$~\cite{chang2013b,chang2015}. Therefore, the single pair of CESs in these systems provides a natural platform for TSC with a 1D spin helical channel. Practically, such a 1D channel can be fabricated as a nanowire~\cite{chen2018} or at the local gate boundary~\cite{zeng2018,lake2018}.

\begin{figure}[t]
\begin{center}
\includegraphics[width=3.3in]{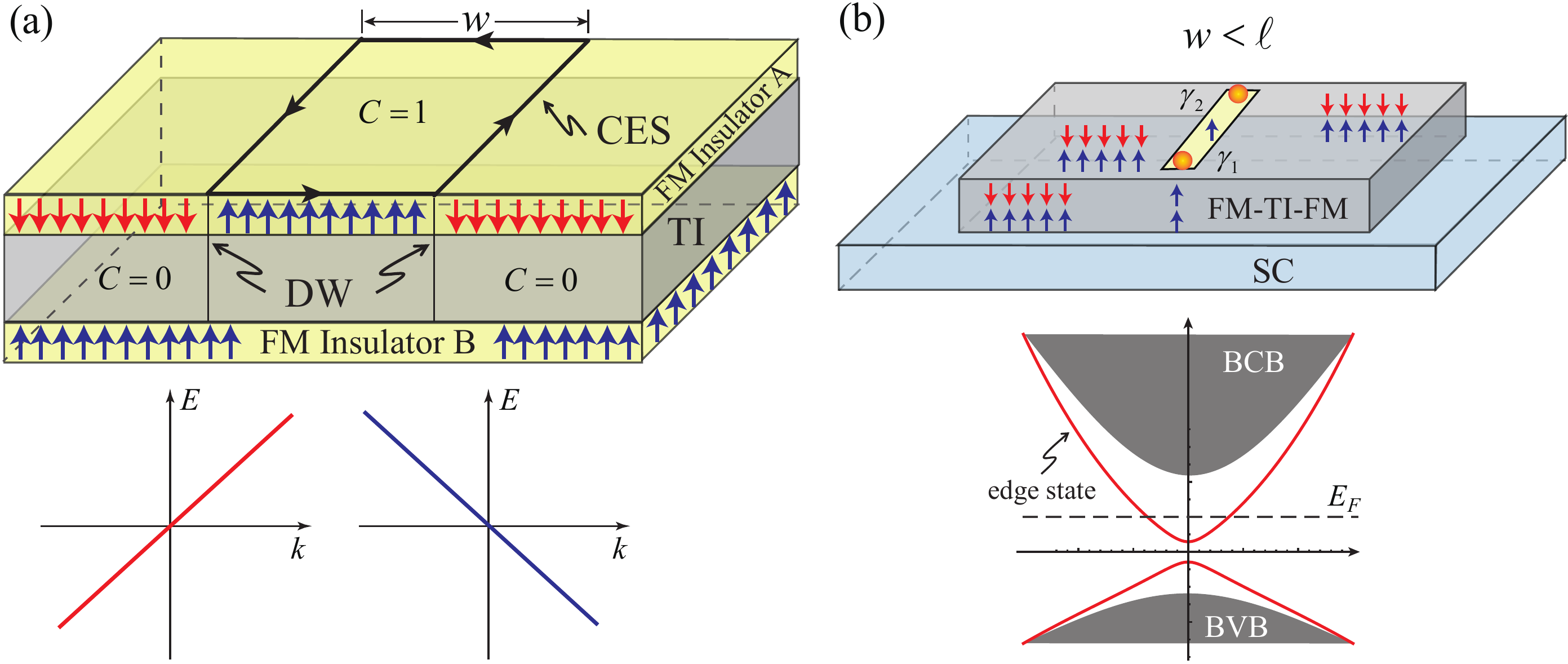}
\end{center}
\caption{MZMs created by magnetic domain writing by MFM in a magnetic TI. (a)	Illustration of a single CES at the magnetic DW between QAH ($C=1$) and axion insulator ($C=0$) in a FM-TI-FM heterostructure. The schematics of counterpropagating CES band structure are in bottom row. (b) The hybrid FM-TI-FM-SC device. A 1D spin-helical channel at the DW stripe in proximity with a conventional superconductor will become a TSC, which gives rise to two end MZMs $\gamma_1$ and $\gamma_2$. The dashed line indicates the Fermi level $E_F$; BCB (BVB), bulk conduction (valence) band.}
\label{fig1}
\end{figure}

Here, in contrast to previous approaches, we propose to realize the 1D helical channel at DWs between up and down magnetic domains in FM-TI-FM heterostructure as shown in Fig.~\ref{fig1}. The FM insulators A and B have different coercivities $H^c_1$ and $H^c_2$, respectively. Assume that both FM A and B have an out-of-plane magnetic easy axis, and the same sign of the exchange coupling parameter to the TI surface states. When A and B have parallel magnetization, the system is in a QAH state with the Chern number ($C=+1$ or $-1$) depending on the magnetization direction ($M>0$ or $M<0$)~\cite{wang2014a,mogi2015,mogi2017}. When A and B have antiparallel magnetization, the system is an axion insulator with $C=0$~\cite{mogi2017,xiao2018}. A gapless CES lies at the DW between these two states as shown in Fig.~\ref{fig1}(a), where the propagating direction of CES is along $\nabla M(\mathbf{r})\times\hat{\mathbf{n}}$, $\hat{\mathbf{n}}$ is the surface normal. The two counterpropagating channels at opposite DWs begin to hybridize when the distance between them $w$ is smaller than the CES width $\ell$. With Fermi level crossing the 1D spin helical band, an effective $p$-wave pairing is induced when it is proximity coupled to an $s$-wave superconductor.

One advantage is that the size, position and shape of the 1D channel can be manipulated via domain control by external fields~\cite{mellnik2014,yasuda2017a,yasuda2017b,rosen2017}. Recently, a technique based on magnetic force microscopy (MFM) has been used to write various domain patterns in magnetic TIs~\cite{yasuda2017b}. The stray field $H_{\text{stray}}$ from the MFM tip could reverse the magnetization, which decays exponentially away from the tip along $z$ direction. To ensure the domain writing on top FM A but not affecting bottom FM B, we need $H^c_1<H_{\text{stray}}<H^c_2$. Experimentally, one can choose the candidate FM materials Cr$_x$(Bi$_{1-y}$Sb$_y$)$_{2-x}$Te$_3$ (CBST) and V$_x$(Bi$_{1-y}$Sb$_y$)$_{2-x}$Te$_3$ (VBST) for A and B, respectively. Both of them are FM insulators with an out-of-plane easy axis, and have a good lattice match with the Bi$_2$Te$_3$ family materials. CBST with $0.2<x<0.4$ has $T_c=40$-$90$~K and $H^c_1\sim0.1$~T~\cite{ou2016}. VBST with $0.1<x<0.3$ has $T_c=20$-$100$~K and $H^c_2\sim1.0$~T~\cite{chang2015}. Moreover, modulation doping in this magnetic TI heterostructure enhance the magnetically induced mass gap from the magnetic proximity effect and to suppress the doping-induced disorder in the surface-state conduction. This leads us to design the device in Fig.~\ref{fig1}(b).

\emph{Model.} Now we turn to the TI film with surface magnetization and superconducting proximity. The low energy physics of the system is described by the Dirac-type surface states only, for the bulk states are gapped. The generic form of the 2D effective Hamiltonian is
\begin{eqnarray}\label{model0}
\mathcal{H}_0(\vec{k})
   &=&v_F k_y\sigma_1\otimes\tau_3-v_F k_x\sigma_2\otimes\tau_3+m(\vec{k})1\otimes\tau_1
   \nonumber\\
   &&+\lambda_s(x,y)\sigma_3\otimes1+\lambda_a(x,y)\sigma_3\otimes\tau_3,
\end{eqnarray}
with the basis of $\psi_{\vec{k}}=(c_{t\uparrow},c_{t\downarrow},c_{b\uparrow},c_{b\downarrow})^T$, where $t$ and $b$ denote the top and bottom surface states, and $\uparrow$ and $\downarrow$ represent the spin up and down states, respectively. $\vec{k}=(k_x,k_y)$. $\sigma_i$ and $\tau_i$ ($i=1,2,3$)
are Pauli matrices acting on spin and layer, respectively. $v_F$ is the Fermi velocity. $m(\vec{k})=m_0+m_1(k_x^2+k_y^2)$, describes the tunneling effect between the top and bottom surface states. The last two terms describe the Zeeman-type spin splitting for surface states induced by the FM exchange couplings along $z$ axis from FM A and B, where $\lambda_s=(\lambda_t+\lambda_b)/2$ is the parallel Zeeman field and $\lambda_a=(\lambda_t-\lambda_b)/2$ is the staggered Zeeman field. $\lambda_j$ ($j=t,b$) are exchange field on top and bottom surface, respectively. In a simple case for uniform $\lambda_j$ with $|\lambda_j|-m_0>0$, the system is a QAH insulator with $C=\lambda_t/|\lambda_t|$ when magnetization is parallel $\lambda_t\lambda_b>0$, and it is an axion insulator with $C=0$ when magnetization is antiparallel $\lambda_t\lambda_b<0$. $\lambda_j$ is nonuniform at a magnetic DW and can be modeled as $\lambda(x,y)=\lambda_0\tanh(2x/l)$, where $l$ is the DW width. For the device in Fig.~\ref{fig1}(b), $\lambda_b(x,y)=\lambda_0$ and $\lambda_t(x,y)=\lambda_1\left[\tanh(2x/l)-\tanh(2(x-w)/l)-1\right]$, here for simplicity we assume $\lambda_0=\lambda_1>0$. The 2D bulk gap is $2(\lambda_0-|m_0|)$. Now in proximity to an $s$-wave superconductor, a finite pairing amplitude is induced in the magnetic TI system. The BdG Hamiltonian becomes $H_\text{BdG}=\sum_{\vec{k}}\Psi^\dag_{\vec{k}}\mathcal{H}_{\text{BdG}}\Psi_{\vec{k}}/2$, with $\Psi_{\vec{k}}=(\psi_{\vec{k}}^T,\psi^\dag_{-\vec{k}})^T$ and
\begin{eqnarray}
\mathcal{H}_{\text{BdG}} &=&
\begin{pmatrix}
\mathcal{H}_0(\vec{k})-\mu & \Delta(\vec{k})\\
\Delta^{\dag}(\vec{k}) & -\mathcal{H}^*_0(-\vec{k})+\mu
\end{pmatrix}\label{BdG},
\\
\Delta(\vec{k}) &=&
\begin{pmatrix}
i\Delta_1\sigma_2 & 0\\
0 & i\Delta_2\sigma_2
\end{pmatrix}.\nonumber
\end{eqnarray}
Here $\mu$ is chemical potential, $\Delta_1$ and $\Delta_2$ are pairing gap functions on top and bottom surface state, respectively.

First, we compare the energy scale of the parameters in the BdG Hamiltonian. $m_0\sim0$-$50$~meV depends on film thickness~\cite{zhang2010}. $\lambda\sim30$-$100$~meV is tunable by changing the magnetic ion doping concentration~\cite{ou2016}. To ensure stable QAH state, $\lambda\gg m_0$. $\Delta\sim0.5$~meV is proximity from an $s$-wave supercondutor~\cite{wangmx2012}. Thus $\lambda\gg m_0,\Delta$ in general. As long as $\mu$ is in the bulk gap, the finite $\Delta$ will not change the bulk topological property of the parent magnetic TI. Namely, $C=1$ QAH becomes topological equivalent $N=2$ chiral TSC, and $C=0$ axion insulator becomes $N=0$ trivial SC~\cite{wang2015c}. Therefore, we can study the low energy edge theory by projection Eq.~(\ref{BdG}) to the 1D edge of DWs. We replace $k_x\rightarrow-i\partial_x$ and
decompose the Hamiltonian as $\mathcal{H}_{\text{BdG}}=H_0+H_1$, in which
\begin{eqnarray}
H_0(-i\partial_x,k_y) &=& v_Fk_y\sigma_1\tau_3+ iv_F\sigma_2\tau_3\zeta_3\partial_x
\nonumber
\\
&&+\lambda_s(x)\sigma_3\zeta_3+\lambda_a(x)\sigma_3\tau_3\zeta_3,
\end{eqnarray}
where $\zeta_{3}$ is the Pauli matrice in Nambu space. We solve $H_0$ first and treat $H_1$ as a perturbation, which is justified since $\Delta$ and $m_0$ are relatively small.

\begin{figure}[t]
\begin{center}
\includegraphics[width=2.3in]{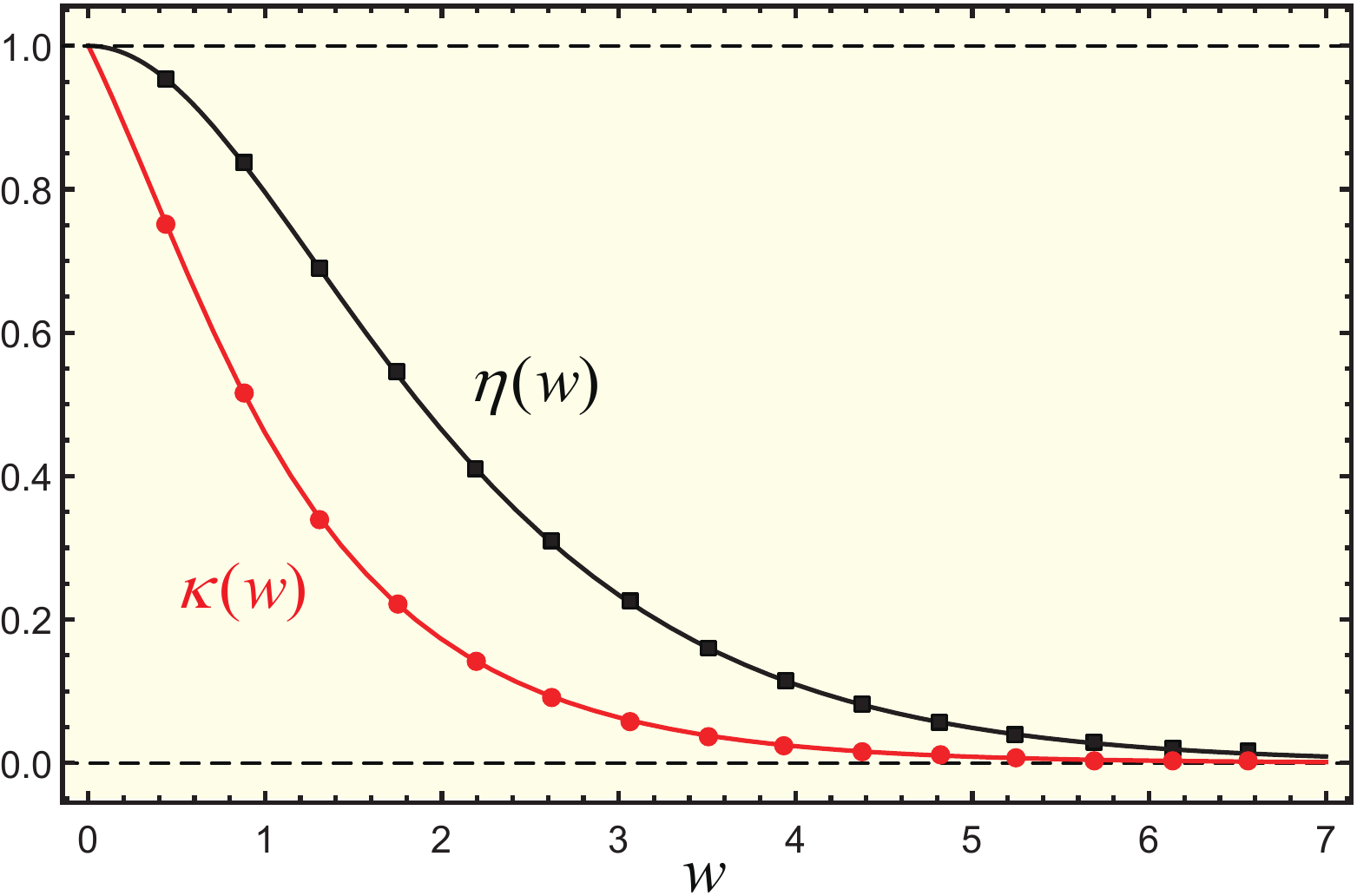}
\end{center}
\caption{The finite size effect of CESs. The $\kappa(w)$ and $\eta(w)$ as a function of $w$, where we set $l=1$.}
\label{fig2}
\end{figure}

Next, we solve the eigenequation $H_0\varphi_{j}(x)=E\varphi_{j}(x)$ at each DW with open boundary condition $\varphi_{j}(-\infty)=\varphi_{j}(\infty)=0$, and find four bound state solutions with the forms $\varphi_{1,3}(x)=\phi(x)e^{ik_yy}\chi_{1,3}$, $\varphi_{2,4}(x)=\phi(x-w)e^{ik_yy}\chi_{2,4}$, $\phi(x)\equiv N\cosh^{-l/2}(2x/l)$. Here $N$ is normalization factor, $\chi_{1,3}=|\sigma_1=+1\rangle\otimes|t\rangle\otimes|\zeta_3=\pm1\rangle$, and $\chi_{2,4}=|\sigma_1=-1\rangle\otimes|t\rangle\otimes|\zeta_3=\pm1\rangle$. The projection of the bulk model onto the lowest four modes leads to 1D effective Hamiltonian
\begin{equation}\label{1D}
\mathcal{H}_{\text{1D}}(k_y)=v_Fk_y\sigma_3+m_t\sigma_1\zeta_3+\Delta_0\sigma_2\zeta_2-\mu\zeta_3,
\end{equation}
where $m_t=\lambda_0\kappa(w)$ is the hybridization of the two counterpropagating CESs, $\kappa(w)=\int_{-\infty}^{\infty}\phi^*(x)(\tanh(2x/l)-1)\phi(x-w)dx$ is a dimensionless function. $\Delta_0=-\Delta_1\eta(w)$ is the effective pairing gap function of 1D channel, $\eta(w)=\int_{-\infty}^{\infty}\phi^*(x)\phi(x-w)dx$ is also dimensionless. Fig.~\ref{fig2} shows the analytical calculations of $\kappa(w)$ and $\eta(w)$. Both $\kappa(w)$ and $\eta(w)$ tends to 0 as the distance $w\rightarrow\infty$, but $\kappa(w)$ decays faster than $\eta(w)$ as $w$ increases. This can be understood that $\kappa(w)$ is only determined by the wavefunction overlap of these two localized counterpropagating CESs, while $\eta(w)$ depends on the hopping of the CESs to the 2D itinerant electrons in superconductors. The excitation spectrum is $E(k_y)=\pm\sqrt{\Delta_0^2+(\sqrt{v_F^2k_y^2+m_t^2}\pm\mu)^2}$, which only vanishes when $\Delta_0=k_y=0$ and $|\mu|=m_t$. For $\mu\gg\Delta_0,m_t$, the low energy spectrum resembles that of a 1D $p$-wave superconductor. Regularizing Eq.~(\ref{1D}) into a lattice model, the $\mathcal{Z}_2$ invariant is evaluated as $\nu=\text{sgn}[\text{Pf}(\widetilde{H}_{\text{1D}}(0))\times\text{Pf}(\widetilde{H}_{\text{1D}}(\pi))]$, where $\text{Pf}$ denotes the Pfaffian number, $\widetilde{H}_{\text{1D}}$ is the skew-symmetrized of $\mathcal{H}_{\text{1D}}$ in the Majorana basis. It is \emph{always} topologically nontrivial with $\Delta_0\neq0$ as long as $\lambda_0-|m_0|>|\mu|>m_t$. Namely, $\nu=-1$ when the chemical potential is inside the 2D bulk gap but outside the hybridization gap of CESs.

\begin{figure}[t]
\begin{center}
\includegraphics[width=3.4in]{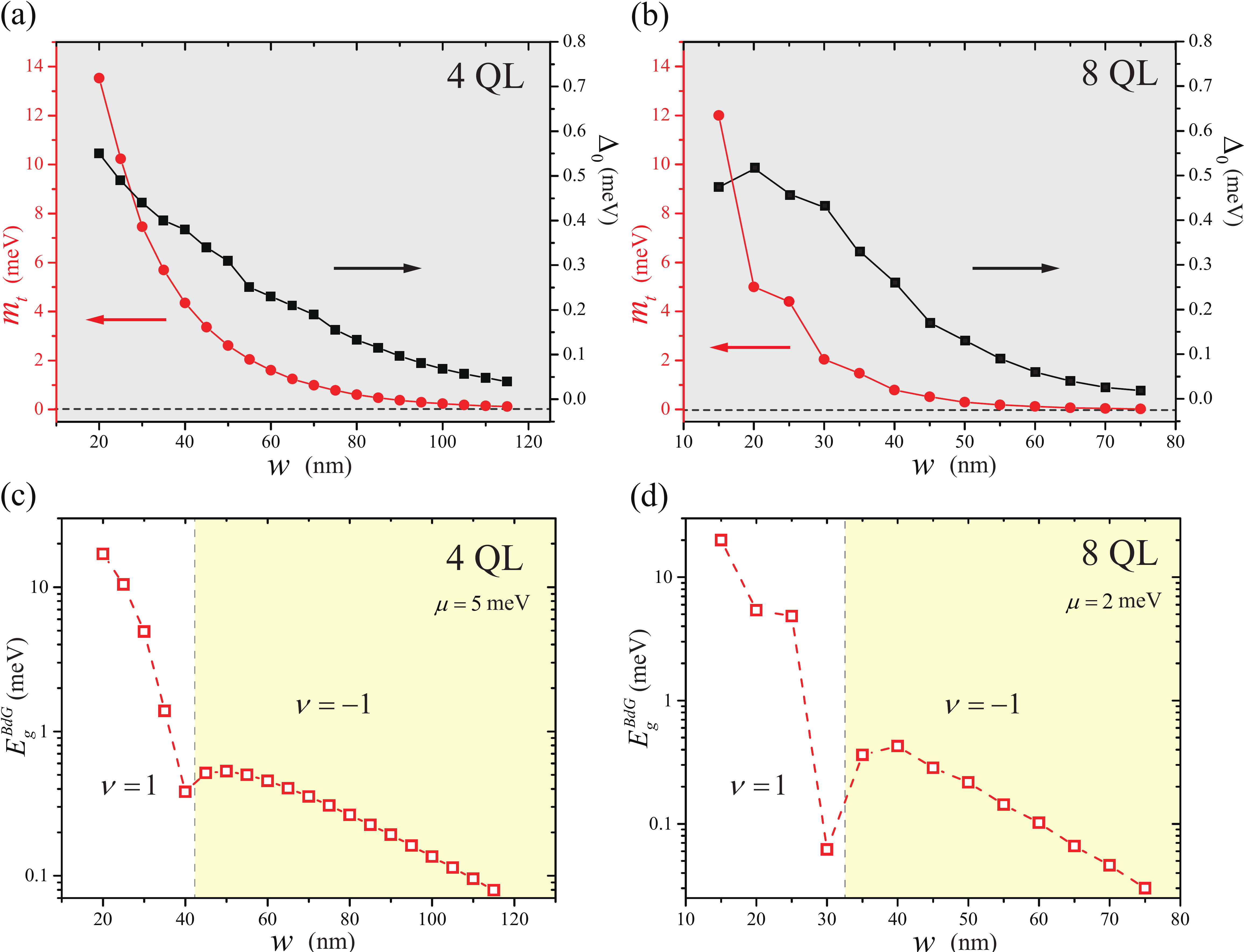}
\end{center}
\caption{(a),(b) The $m_t(w)$ and $\Delta_0(w)$ in 4 and 8 QL TI, respectively, as a function of the DW stripe width $w$. Each QL is about 1 nm thick. The hybridization gap between top and bottom surface states is $m_0\approx15$~meV in 4 QL, while it is negligibly small in 8 QL. (c),(d) The gap of BdG Hamiltonian $E_{g}^{\text{BdG}}$ vs $w$ for $\mu=5$~meV in 4 QL, and $\mu=2$~meV in 8 QL, respectively.}
\label{fig3}
\end{figure}

The above discussion based on the effective model gives us a clear physical picture of the 1D TSC realized at DWs in magnetic TIs, which is generic for any TI system and do not rely on a specific model. For concreteness, to estimate the magnitude of $m_t(w)$ and $\Delta_0(w)$, we adopt the effective Hamiltonian in Ref.~\cite{wang2013a} to describe the low-energy bands of Bi$_2$Te$_3$ family materials, $\mathcal{H}_{\mathrm{3D}}(x,k_y,z)=\varepsilon 1\otimes1+d^1\tau_1\otimes1+d^2\tau_2\otimes\sigma_3+d^3\tau_3\otimes1-\lambda(x,z)\tau_3\otimes\sigma_3+d^5\tau_2\otimes\sigma_2$. Here $\varepsilon(x,k_y,z)=-D_1\partial_z^2-D_2\partial_x^2+D_2k_y^2$, $d^{1,2,3,5}(x,k_y,z)=(-iA_2\partial_x, A_2k_y, B_0-B_1\partial_z^2-B_2\partial_x^2+B_2k_y^2, iA_1\partial_z)$, and $\lambda(x,z)$ is the $x,z$-dependent exchange field. We then discretize it into a tight-binding model along both $z$-axis between neighboring quintuple layers (QL) and $x$-axis from $\mathcal{H}_{\text{3D}}$, and assume in bottom layer $\lambda(x,-d/2)=\lambda_0$, in top layer $\lambda_0(x<0,d/2)=\lambda(x>w,d/2)=-\lambda_0$ and $\lambda_0(0<x<w,d/2)=\lambda_0$, and zero elsewhere. We further assume the pairing gap function $\Delta(z)=\Delta_1$ in the top layer, and zero elsewhere. The total length along $x$ axis is chosen as $5w$, and $d$ is the film thickness along $z$ axis. Fig.~\ref{fig3}(a) and~\ref{fig3}(b) shows the numerical calculations of parameters $(m_t(w),\Delta_0(w))$ for thin films of 4 and 8 QL, respectively, where we set $\Delta_1=1.5$~meV and a typical surface exchange field $\lambda_0=30$~meV~\cite{lee2015}. All the other parameters are taken from Ref.~\cite{zhang2009} for (Bi$_{0.2}$Sb$_{0.8}$)$_2$Te$_3$. As is consistent with the analytical results in Fig.~\ref{fig2}, $m_t(w)/\lambda_0$ decays faster than $\Delta_0(w)/\Delta_1$ as $w$ increases. With a fixed chemical potential, this leads to a \emph{wide} range of DW width to be topologically nontrivial as shown in Fig.~\ref{fig3}(c) and~\ref{fig3}(d).  Here we mention that the topological phase transition is accompanied by BdG gap closing, which is not shown in Fig.~\ref{fig3} due to discrete in plane lattice size. The second advantage of this system is that the topological regime is larger compared to that of semiconductor nanowire system. Take $w=50$~nm in 4~QL for example, with a large topological regime of about $12.4$~meV in terms of $\mu$, one still has sizable superconducting gap of about $\Delta_0\approx0.34$~meV.

\emph{Experimental feasibility.} Now we discuss the feasibility of the proposals. Experimentally, to obtain MZM at the end of DW stripe in TIs, one must fulfill the following requirements. First, finely tune the Fermi level into the magnetically induced surface gap and keep the bulk truly insulating, but outside the hybridization gap of CESs. Second, a good proximity effect between the conventional superconductor and magnetic TI heterostructure is necessary. Third, the DW stripe is much longer than twice the localized length of MZM. Recently experimental progresses have already shown in magnetic TIs the good chemical potential tunability by an external gate~\cite{yasuda2017b}, and superconducting proximity effect with Nb~\cite{he2017}, fulfilling the first two conditions above, providing a good platform to observe the MZM. For finite pairing, the localized length of the end MZM can be estimated by $v_F/2\Delta_0$, with $v_F=2.0$~eV{\AA}, the localization length is about $0.4$~$\mu$m. Therefore one expects to see a MZM at each end of DW stripe with length larger than $1$~$\mu$m, which can be easily verified by scanning tunneling microscope (STM).

From Fig.~\ref{fig3}, we learn that the stripe width of topologically nontrivial DWs depends on 2D bulk gap. However, there is unavoidable spatial fluctuation of the exchange field due to the inhomogeneity of the Cr concentration~\cite{lee2015}, which reduces the effective size of the bulk gap. The smaller effective bulk gap results in the larger CES width, and a larger stripe width for topologically nontrivial DWs. Take 4 QL for an estimation, with effective $\lambda_0=18$~meV and $\mu=1$~meV, the optimal width of DW stripe is $90~\text{nm}<w<250$~nm, which is within the $s$-wave superconducting coherence length of about several hundreds nm. Here the upper bound is limited by the energy resolution of STM about $0.1$~meV. Moreover, the localized in-gap states due to the doping, vacancies and defects in magnetic TI are well distinguishable from MZMs because of different energy and position dependence. The top-bottom asymmetry will further reduce the bulk gap~\cite{wang2015a}, and should be avoided.

\begin{figure}[t]
\begin{center}
\includegraphics[width=3.3in]{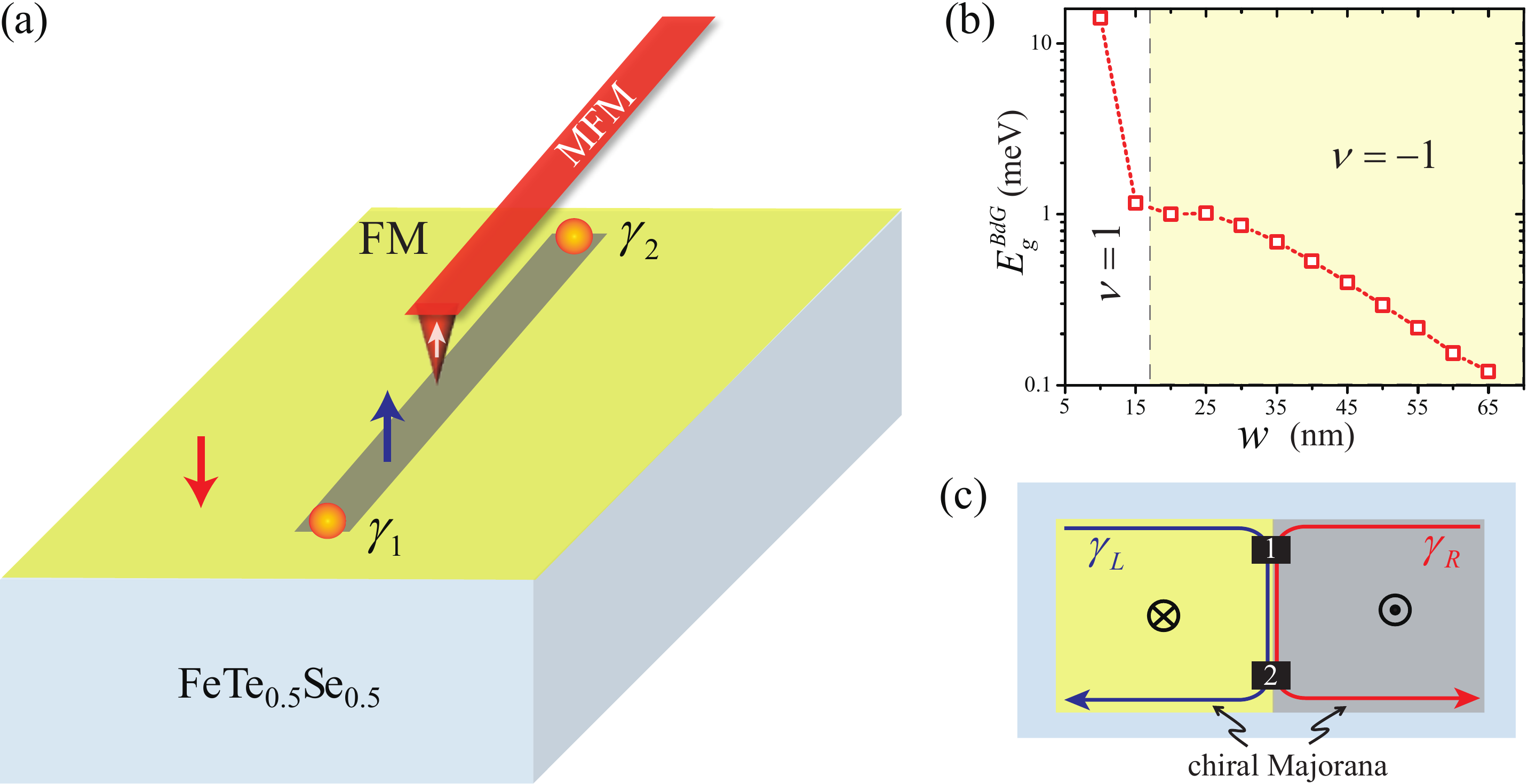}
\end{center}
\caption{(a) 1D effective TSC with end MZMs realized at magnetic DWs stripe on superconducting topological surface state, and possible application to FeTe$_{0.5}$Se$_{0.5}$. (b) The width of topological nontrivial DW stripe with $\lambda=10$~meV. (c) Two chiral Majorana edge fermions $\gamma_L$, $\gamma_R$ of opposite chirality (indicated by blue and red arrows) at two opposite magnetic domain boundaries on FeTe$_{0.5}$Se$_{0.5}$ surface, mix into a one-way charge conducting channel at DW. $\bigotimes$ and $\bigodot$ denotes the pointing in and out magnetization direction.}
\label{fig4}
\end{figure}

\emph{FeTe$_{0.5}$Se$_{0.5}$} The MZMs discussed in the above proposal could be further extended to DWs on superconducting Dirac surface states in FeTe$_{0.5}$Se$_{0.5}$~\cite{wangzj2015,wu2016,xu2016,zhang2018}. Superconductor FeTe$_{1-x}$Se$_x$ ($x\sim0.5$) is predicted to be topologically nontrivial with a single Dirac cone on the surface. Recently, a spin-resolved angle-resolved photoemission spectroscopy experiment has confirmed that the spin-helical surface electrons at the Fermi level open an $s$-wave superconducting gap of about $1.8$~meV below $T_c\sim14.5$~K~\cite{zhang2018}. Such 2D superconducting surface state is a TSC which resembles the spinless $p+ip$ superconductor but does not violate time-reversal symmetry. A pair of MZMs appear at the two ends of vortices. Moreover, a chiral Majorana fermion exists at the edge of a magnetic domain deposited on surface. The effective BdG Hamiltonian describing the superconducting surface with surface magnetization at $\Gamma$ point in this system is
\begin{equation}
\mathcal{H}_b = vk_y\sigma_1-vk_x\sigma_2\zeta_3+\lambda(x,y)\sigma_3\zeta_3-\Delta\sigma_2\zeta_2-\mu\zeta_3,
\end{equation}
where $\lambda(x,y)$ is the Zeeman field induced by proximity to FM insulator. This 2D model is similar to Eq.~(\ref{BdG}) in the $m(\vec{k})\rightarrow0$ limit. Therefore, one expect to get the same effective 1D TSC model at DW stripe as Eq.~(\ref{1D}). The proposed device on FeTe$_{0.5}$Se$_{0.5}$ surface is shown in Fig.~\ref{fig4}(a), where the magnetic proximity from FM insulator will not destroy the bulk superconductivity in FeTe$_{0.5}$Se$_{0.5}$. For an estimation of the DW stripe width, we take $\mu=6$~meV,  $v=1.4$~eV{\AA}~\cite{zhang2018}, and choose $\lambda=10$~meV, the stripe is topologically nontrivial when the width is larger than $20$~nm as shown in Fig.~\ref{fig4}(b). The nontrivial superconducting gap is less than 1~meV, which is easily accessible in STM and within the bulk gap size of 2.5~meV for the hole band at $\Gamma$ point and 4.2~meV for the electron band at $M$ point. The lower bound for topologically nontrivial stripe width becomes smaller when $\lambda$ increases. The candidate FM material can be chosen as CBST. Besides of MZM, one can also get a one-way conducting channel for electric charge at DW, where two counterpropagating chiral Majorana edge fermions mix as shown in Fig.~\ref{fig4}(c). Such mixing of opposite chirality Majorana edge fermions is hard to achieve in the $N=1$ chiral TSC in magnetic TI with an external magnetic field~\cite{he2017}. Unlike the mode mixing in bipolar integer quantum hall junctions~\cite{williams2007}, the single conducting channel here leads to anomalous nonlocal conductance $\sigma_{12}$ and resistance $R_{12}$~\cite{serban2010}.

The unitary operations in topological quantum computing are from the non-abelian braiding of MZMs, which requires $T$ junction structure in wire network~\cite{alicea2011}. Such $T$ junctions network can be written and erased by MFM tip. With a tunable scanning rate fulfilling the adiabatic exchange condition of MZMs, we expect that the realization of Pauli $Z$ gate and controlled-NOT gate is promising.

\begin{acknowledgments}
We acknowledge Tong Zhang for valuable discussions. This work is supported by the Natural Science Foundation of China through Grant No.~11774065; the National Key Research Program of China under Grant No.~2016YFA0300703; the Natural Science Foundation of Shanghai under Grant No.~17ZR1442500; the National Thousand-Young-Talents Program; the Open Research Fund Program of the State Key Laboratory of Low-Dimensional Quantum Physics, through Contract No.~KF201606; and by Fudan University Initiative Scientific Research Program.
\end{acknowledgments}

\end{document}